# Effect of doping on the intersubband absorption in Si- and Ge-doped GaN/AlN heterostructures

A Ajay[1,2], C B Lim [1,2], D A Browne[1,2], J Polaczynski[1,3], E Bellet-Amalric[1,2], J Bleuse[1,2], M I den Hertog[1,3], E Monroy[1,2]

[1]Université Grenoble-Alpes, 38000 Grenoble, France.
[2]CEA-Grenoble, INAC-PHELIQS, 17 av. des Martyrs, 38000 Grenoble, France.
[3]CNRS-Institute Néel, 25 av. des Martyrs, 38000 Grenoble, France.

**ABSTRACT**

In this paper, we study band-to-band and intersubband characteristics of Si- and Ge-doped GaN/AlN heterostructures (planar and nanowires) structurally designed to absorb in the short-wavelength infrared region, particularly at 1.55 µm. Regarding the band-to-band properties, we discuss the variation of the screening of the internal electric field by free carriers, as a function of the doping density and well/nanodisk size. We observe that nanowire heterostructures consistently present longer photoluminescence decay times than their planar counterparts, which supports the existence of an in-plane piezoelectric field associated to the shear component of the strain tensor in the nanowire geometry. Regarding the intersubband characteristics, we report absorption covering 1.45 µm to 1.75 µm using Ge-doped quantum wells, with comparable performance to Si-doped planar heterostructures. We also report similar intersubband absorption in Si- and Ge-doped nanowire heterostructures indicating that the choice of dopant is not an intrinsic barrier for observing intersubband phenomena. The spectral shift of the intersubband absorption as a function of the doping concentration due to many body effects confirms that Si and Ge efficiently dope GaN/AlN nanowire heterostructures.

# 1. Introduction

The intersubband (ISB) technology has gone a long way to become a leading optoelectronics platform since the first observation of ISB transitions using quantum confinement in GaAs/Al$_x$Ga$_{1-x}$As multi-quantum wells by Esaki and Sasaki [1]. GaAs-based ISB detectors are the main choice for the long-wavelength (LWIR, 8-14 µm) or very-long-wavelength (VLWIR, 14-30 µm) infrared windows [2]. For the short-wavelength infrared (SWIR, 1–3 µm), band-to-band mercury-cadmium telluride devices is most often chosen in the 1.6-3 µm range, whereas the 1-1.6 µm range (including the 1.3-1.5 µm fiber optic communication window) is covered by InGaAs band-to-band devices. There, ISB devices would present interesting advantages in terms of bandwidth, but ISB transitions in the SWIR require larger band offsets than those of GaAs-based semiconductors. GaN/AlN heterostructures are a promising system for the development of a SWIR ISB technology [3], owing to their large conduction band offset of about 1.8 eV [4–6] and sub-picosecond ISB relaxation time [7–9]. There are various reports on GaN-based ISB devices in the literature, namely all-optical switches, electro-optical modulators, and infrared photodetectors, including quantum cascade detectors [3].

The n-type doping level is a critical parameter for the performance of ISB devices such as photodetectors or modulators, as the first energy level in the conduction band has to be populated in order to allow transition to higher levels. Thus, the study of high doping levels in GaN-based heterostructures without causing structural degradation of the material is crucial for the development of ISB devices. So far, all GaN-based ISB studies involving planar layers have used Si as n-type dopant. However, it is generally accepted that Si introduces tensile strain in the GaN layers, and doping with Si in excess of $10^{19}$ cm$^{-3}$ can lead to surface roughening and crack propagation [10]. Ge has been introduced as an alternative n-type dopant for GaN, and high doping levels (>$10^{20}$ cm$^{-3}$) have been demonstrated without causing structural degradation of the layers [11–15]. Yet Ge remains unexplored as n-type dopant for GaN-based ISB devices.

On the other hand, using nanowires instead of planar structures as active media in ISB devices can lead to improved performance due to their lower electrical cross-section. Nanowires can have a diameter smaller than the detected/emitted wavelength, and present a large dielectric mismatch with their surroundings. Such features allow engineering the refractive index while maintaining the absorption characteristics of the bulk [16–19]. Enhanced absorbance at certain resonance wavelengths, greater than that predicted by Lambert-Beer law for equivalent thickness of thin films, have also been reported in nanowires [20–22]. Additionally, their large surface-to-volume ratio allows misfit strain to be elastically released, hence expanding the design possibilities of a defect-free active region. Furthermore, three-dimensional confinement of carriers in nanowire heterostructures might open new possibilities of control of the carrier relaxation time [23,24]. However, the field of nanowire ISB transitions is still in the nascent phase. Tanaka et al. [25] reported ISB absorption centered at 1.77 µm (= 0.7 eV) with full width at half maximum (FWHM) of about 230 meV in a GaN/AlN (1 nm/2.7 nm) periodic heterostructure in GaN nanowires with the AlN barriers doped with Si at 2×$10^{19}$ cm$^{-3}$. In this configuration, it is assumed that the electrons from the donor levels in the AlN barriers should be transferred to the GaN nanodisks. Studies in GaN/AlN planar heterostructures indicate an improvement of the ISB absorption linewidth if the doping is performed directly in the GaN wells [26], but today, to the best of our knowledge, there is no report on intersubband absorption in GaN/AlN nanowire heterostructures with Si-doped GaN nanodisks. Difficulties have been attributed to the fact that Si tends to degrade nanowire morphology and migrate towards the nanowire sidewalls, resulting in inefficient doping [27]. Replacing Si by Ge, Beeler et al, [28] utilized GaN:Ge/AlN (4-8 nm /4 nm) heterostructures on GaN nanowires to observe ISB absorption targeting the SWIR. The transitions were broad though, with a FWHM of ≈ 400 meV. Recently, we have announced that, using the adequate growth conditions, it is possible to observe ISB in GaN/AlN nanowire heterostructures with Si-doped GaN sections [29]. In view of these results, further studies are required to conclude on the real impact of the dopant choice and concentration.

In this paper we assess the optical properties (band-to-band and ISB) of n-type doped GaN/AlN superlattices in both planar and nanowire geometries, using the same concentration of Ge and Si dopants in the GaN sections in both geometries, and varying the concentrations in the $1×10^{18}$ to $3×10^{20}$ cm$^{-3}$ range. The structures are designed to display ISB transitions in the SWIR, around 1.55 µm. This manuscript reports, to our knowledge, the first study of ISB absorption in GaN quantum wells with Ge doping. We demonstrate a spectral shift of the PL and ISB transition as a function of doping due to may-body effects, and correlate it with theoretical calculations. Furthermore, we report the first observation of many-body effects on the ISB absorption of Si-doped GaN/AlN nanowire heterostructures. This observation is particularly relevant since it proves that, with our growth conditions, Si can efficiently dope the GaN nanowires without material degradation. Finally, we show that nanowire heterostructures consistently present longer PL decay times than their planar counterparts with identical nanodisk/well size and doping level, which confirms the existence of an in-plane electric field in the nanowire geometry.

## 2. Methods

The structural quality of the samples was analyzed by high-resolution x-ray diffraction (HR-XRD) using a Rigaku SmartLab x-ray diffractometer. For planar layers, we utilized a 4 bounce Ge(220) monochromator and a 0.114° long plate collimator in front of the detector, and for the nanowires, an open detector with Si compensator. The morphology of the as-grown nanowire ensemble was studied by field-emission scanning electron microscopy (SEM) using a Zeiss Ultra 55 or a Zeiss 55 microscope. Detailed structural studies were conducted using high-angle annular dark-field scanning transmission electron microscopy (HAADF-STEM) performed on a probe corrected FEI Titan Themis microscope operated at 200 kV.

Low-temperature ($T$ = 5 K) photoluminescence (PL) spectra were obtained by excitation with a Nd-YAG laser ($\lambda$ = 266 nm, pulse width = 0.5 ns, repetition rate = 8 kHz), with an optical power around 100 µW focused on a spot with a diameter of ≈ 100 µm. The emission from the sample was collected by a Jobin Yvon HR460 monochromator equipped with an ultraviolet-enhanced charge-coupled device (CCD) camera. For time-resolved PL measurements, excitation was provided by a frequency-tripled Ti:sapphire laser ($\lambda$ = 270 nm) with pulse width of 200 fs. The laser was augmented with a cavity damper section with a base pulse repetition rate of 54 MHz, allowing the period between pulses to be varied from 20 ns to 500 ns. The incident beam had an excitation power of 250 µW. The luminescence was dispersed using a Jobin Yvon Triax320 monochromator and detected by a Hamamatsu C-10910 streak camera. Room-temperature Fourier transform infrared (FTIR) spectroscopy was used to probe the ISB absorption using a halogen lamp, a $CaF_2$ beam splitter, and a mercury-cadmium-telluride detector incorporated into a Bruker V70v spectrometer. Transmission in nanowire samples was measured at nearly grazing angles of 5°. In the case of planar samples, they were polished at 45° to form multipass waveguides allowing 4–5 interactions with the active region. An infrared polarizer was used to discern between the transverse-electric (TE) and transverse-magnetic (TM) polarized light.

To understand the band diagram and energy states of the planar and nanowire structures, one-dimensional and three-dimensional (3D) calculations were performed using Nextnano[3] Schrödinger-Poisson equation solver [30]. In the case of planar structures, we performed one-dimensional calculations assuming that the strain configuration in the superlattice is such that the in-plane lattice parameter *a* is equal to the one expected for relaxed AlGaN with the average Al mole fraction of the superlattice. This approximation is based on the previous study of the strain distribution in GaN/AlN superlattices on various substrates [31]. In the case of the nanowires, three dimensional simulations were performed. The nanowire was defined as a regular hexagonal prism with minor radii being 30 nm. Along the length, it consisted of a 150-nm-long base GaN section, followed by a heterostructure of 14 periods of AlN (3 nm) and GaN (1.7 nm), and a 50-nm-thick GaN cap. The GaN base and the GaN/AlN superlattice were laterally surrounded by a 5-nm-thick AlN shell, and the whole structure enclosed in a rectangular prism of air, permitting elastic deformation.

## 3. Sample structure

Planar and nanowire samples with an active superlattice consisting of 30 repetitions of 2 nm GaN / 3 nm AlN were synthesized by plasma-assisted molecular-beam epitaxy (PAMBE). Controlled doping of the GaN sections of the superlattices at concentrations of $3\times10^{19}$ cm$^{-3}$, $1\times10^{20}$ cm$^{-3}$, and $3\times10^{20}$ cm$^{-3}$ was achieved using Ge and Si effusion cells. The calibration of the dopant density was performed by Hall effect characterization using the Van der Pauw method on GaN planar layers doped with various concentrations of Si and Ge. The list of the samples under study, with their structure, dopant nature/concentration and other properties are provided in Table I.

Planar heterostructures were grown on 1-µm-thick (0001)-oriented AlN-on-sapphire templates under metal-rich conditions, as described in ref. [26]. The growth rate was 400 nm/h and the substrate temperature was $T_S$ = 720°C. The growth started with the deposition of a 340-nm-thick GaN layer, prior to the AlN/GaN superlattice, which was finally capped with 30 nm of AlN. The higher polarization modulus of the AlN cap with respect to the superlattice results in a positive charge sheet which pins the Fermi level at the conduction band at the superlattice/cap heterointerface.

Self-assembled (000–1)-oriented GaN nanowires were synthesized on floating-zone Si(111) substrates using nitrogen-rich conditions. The growth started with the deposition of an AlN buffer using a two-step procedure, as described elsewhere [29]. Then, a 700-nm-long GaN base was grown at a substrate temperature $T_S$ = 810°C and with a growth rate of 330 nm/h. For the fabrication of the GaN/AlN superlattice, the GaN wells were grown using the same nitrogen-rich conditions that apply to the GaN base, and the AlN sections were grown at the stoichiometry. Finally, the superlattice was capped with 30 nm of GaN. Note that the crystal orientation of the nanowire samples is inverted with respect to the planar samples [32]; therefore, for the nanowires, it is a GaN cap, with lower polarization modulus with respect to the superlattice, which ensures the pinning of the Fermi level at the conduction band at the superlattice/cap heterointerface. The choice of pinning the Fermi level at the topmost heterointerface in both planar and nanowire heterostructures aims at reducing the sensitivity of the structures to {0001} surface states.

The periods of the GaN/AlN superlattices extracted from the intersatellite angular distance of the θ-2θ HRXRD scans around the (0002) reflections are listed in Table I for all the samples under study. There was no distinguishable difference between Si- and Ge-doped samples from the θ-2θ scans [29]. Regarding the nanowire samples, figure 1(a) presents an SEM image of NG1 showing well separated bundles of 1-2 nanowires, as commonly observed for GaN wires grown by PAMBE [33]. The coalescence of the wires, if present, occurs early during the growth and is hence widely distant from the active region. The diameter of the bundle is in the range of 40-80 nm (depending on the number of nanowires present). The schematic structure of a single wire is depicted in figure 1(b). HAADF-STEM images of samples NS1 and NG2 (2 nanowires per sample), similar to figures 1(c) and (d), were used for a statistical analysis of individual GaN quantum wells and AlN barriers to conclude that the superlattices have an average period consisting of 2.0±0.3 nm of GaN and 2.8±0.3 nm of AlN. The first two-three quantum wells are characterized by a higher irregularity in the thickness, which can deviate up to two monolayers (0.5 nm) from the average. The figure also shows the presence of a ≈ 5-nm-thick AlN shell surrounding the nanowire heterostructures, which is known to introduce a uniaxial compressive strain in the structure [28]. The shell thickness gradually decreases as we move away from the substrate. SEM and HAADF-STEM images did not show any morphological change (nanowire diameter, density or facet orientation) as a function of the doping level, whatever the nature of the dopant, Si or Ge.

## 4. Results and discussion

### a) Planar heterostructures

In a first stage, we analyze the band-to-band behavior of the samples by low-temperature ($T$ = 5 K) PL, as illustrated in figure 2. First, we focus on the behavior of planar layers (figures 2(a) and (c) for Si and Ge doping, respectively), the spectra present a multi-peak structure arising from in-plane monolayer fluctuations of the quantum well thickness [6,26]. For the samples with lower doping level, the emission is located slightly below the GaN bandgap due to the polarization-induced internal electric field, as expected for this well width. For increasing dopant concentration, there is a blueshift in the peak emission, which can be assigned to the screening of the internal electric field by the free carriers [34]. The behavior is very similar for both dopants.

We have compared the magnitude of the screening effect with the result of one-dimensional simulations using the nextnano³ Schrodinger-Poisson-solver. The solid lines in figure 3(a) represent the emission wavelength (or energy) associated with the transition between the lowest electron energy level, $e_1$, in the conduction band and the highest hole level, $h_1$, in the valence band of the quantum wells. For the calculations, we varied the doping level and the thickness of the GaN quantum well, whereas the AlN barrier was chosen to be 3 nm in all the cases (the error in the $e_1$-$h_1$ transition associated with this choice is negligible). The impact of screening of the electric field due to doping is much reduced in smaller quantum wells compared to large quantum wells. This is explained by the fact that free carrier screening effects scale with the ratio between the well thickness and the Debye length [34], the latter defined as $\lambda_D = \sqrt{\varepsilon_r \varepsilon_0 k_B T/(q^2 N_D)}$, where $\varepsilon_r \varepsilon_0$ is the dielectric constant times the vacuum permittivity, $k_B T$ is the thermal energy, $q$ is the electron charge and $N_D$ is the dopant density.

Experimental data are superimposed to the theoretical calculations in figure 3(a). Data are represented by a set of error bars where the horizontal error bar represents the uncertainty in the well thickness and the vertical error bar, the PL peak wavelengths obtained for Si and Ge doping (see Table I for precise values). The experimental blueshift shows the same trend as the theoretical expectations.

To confirm the screening of the electric field by free carriers, we have studied the band-to-band dynamics of the samples using time-resolved PL. After the laser pulse hits the sample, the luminescence presents a red shift of 8-15 nm during the first 10-20 ns. This initial red shift is associated with a non-exponential drop in PL intensity, which is followed by an exponential decay. An example of PL decay transient is illustrated in the inset of figure 4, where the intensity data (normalized to its maximum) corresponds to the maximum intensity registered as a function of time, after the excitation pulse (at t = 0). The spectral red shift (not shown) and non-exponential decay arise from the perturbation in the band structure induced by the large population of electron-hole pairs due to the laser excitation, as previously observed [35,36]. Hence in our analysis we have only utilized the time constant extracted from the exponential decay part, with the results listed in Table I and presented in figure 4. We observe a decrease in the decay time constant with increasing dopant concentration, which confirms the improved electron-wavefunction overlap due to the decreased quantum confined Stark effect in the quantum wells. A similar drop in the decay time constant was observed for both dopants.

Room-temperature FTIR spectroscopy was used to probe the ISB absorption in the heterostructures. In the case of the planar structures, ISB absorption around the 1.55 μm telecommunication wavelength with Si-doped GaN/AlN heterostructures is well documented in the literature [3]. Here, the normalized absorption spectra for TM-polarized light are displayed in figures 5 (a) and (c). We observe Ge-doped heterostructures exhibiting absorption characteristics covering 1.75 to 1.45 μm, associated to the transition from the first to the second electronic level in the quantum wells (e₁ to e₂). The measurements are compared with Si-doped heterostructures using the exact same doping concentration and structure. The multi-peak spectral profile is due to in-plane thickness fluctuations in the quantum wells [6,26]. This structure is spectrally resolved because the linewidth of the transitions is smaller than the energetic difference associated to increasing the quantum well thickness by one monolayer. The blueshift and broadening of the ISB transition with doping is assigned to many-body effects, specifically exchange interactions and plasmon screening or depolarization [37].

We have calculated the $E_{21}$ = e₂–e₁ transition energy of the planar samples with one dimensional calculations using nextnano³ considering a doping density of 1×10$^{17}$ cm$^{-3}$ (table II). Then, following the formulation in [37], we estimated the transition energy corrected for many-body effects:

$$E'_{21} = E_{21}\sqrt{1+\alpha} - E_{ex} \qquad (1)$$

where $\alpha$ is the depolarization coefficient, and $E_{ex}$ is the shift due to the exchange interaction. The excitonic shift and Coulombic attraction shift are expected to be negligible [37]. The calculated values of $\alpha$, $E_{ex}$ and $E'_{21}$ are also presented in table II, together with the experimental measurements, which show an excellent agreement with the $E'_{21}$ calculation.

In summary, we have demonstrated ISB transitions using Ge-doped GaN quantum wells behaving identical to Si-doped samples. Si and Ge are hence assessed to be equally suitable for the doping of ISB devices using a planar architecture.

    b) Nanowire heterostructures

The low temperature PL spectra of the nanowire samples with different doping levels are displayed in figures 2(b) and (d) for Si and Ge doping, respectively. In the spectrum from N0, the GaN stem of the nanowires contributes to a peak emission around 3.44 eV. This spectral location indicates that the stem is significantly strained due to the AlN shell, compared to the unstrained GaN ($\approx$ 3.47 eV). The superlattice-related transition of the undoped nanowire sample (N0) is strongly blue-shifted with respect to the lightly doped planar samples (PS0, PG0). The variation in thickness between the samples can only justify a blueshift of 10 nm, significantly smaller than what is observed (30 nm). The additional blueshift is due to the uniaxial compressive strain along the growth axis imposed by the presence of the AlN shell in nanowire heterostructures, as previously observed and modeled [28,38,39].

For all doped nanowires, we observe a slight blueshift compared to N0, which can be attributed again to the onset of screening of the internal electric field with the introduction of dopants. However, the magnitude of the shift is much reduced in comparison with previous results in Ge-doped GaN/AlN nanowire heterostructures [36,40]. To understand this discrepancy, we have compared the magnitude of the screening effect with the result of 3D simulations using the nextnano$^3$ Schrodinger-Poisson solver. We assumed an AlN barrier thickness of 3 nm and varied the GaN nanodisk thickness and doping level. For each nanodisk thickness, we calculated the 3D strain distribution by minimizing the elastic energy through the application of zero-stress boundary conditions at the surface. We then calculated the band profiles taking into account the piezoelectric fields and assuming a negative charge density of $2\times10^{12}$ cm$^{-2}$ at the air/nanowire sidewall interface. Figure 3 (b) represents the expected emission wavelength associated with the transition between the lowest electron energy level, $e_1$, in the conduction band and the highest hole level, $h_1$, in the valence band of the nanodisks. Note that the calculated transitions are blue-shifted with respect to planar layers (in figure 3 (a)), as previously explained due to compressive strain imposed by the presence of the AlN shell. Similarly to figure 3 (a), emission data from the samples under study are represented by a set of error bars where the horizontal error bar represents the uncertainty in the nanodisk thickness and the vertical error bar, the PL peak wavelengths obtained for Si and Ge doping (precise values compiled in Table I). The PL peak wavelengths from our previous work on Ge-doped GaN/AlN nanowire heterostructures [28] are also displayed in the figure as squares, with the corresponding doping levels indicated in the legend. Owing to larger nanodisks, the PL shift is enhanced in comparison to this work, in agreement with our calculations.

The PL decay time constants measured for the nanowires with different doping levels are displayed in table I and figure 4. The decay time constant decreases with increase in doping, which is consistent with the onset of screening of the internal electric field by free carriers. Furthermore, we observe that the decay time constant in the nanowires is systematically longer than in their planar counterparts, in spite of the slightly smaller well width. The increased decay time constant for the nanowires is explained by the lateral separation of electron and hole wavefunctions due to the lateral electric field induced by the shear component of the strain in the GaN nanodisks, as proposed in [36].

We subsequently carried out room-temperature studies of the polarized infrared transmission in the nanowire heterostructures. We observed an absorption band for TM-polarized light in the Si- and Ge-doped nanowires as displayed in figures 5 (b) and (d), respectively. Similar measurements of the undoped nanowire sample N0 were used as a reference. The absorption corresponds to the transition between the ground electron level of the GaN disks and the first excited electron level associated to confinement along the growth axis (s-$p_z$). The absorption peak lies around 1.39-1.55 µm and 1.37-1.53 µm for Si-doped and Ge-doped nanowire heterostructures NS1 and NG1, respectively. Compared to the previous report of Ge-doped GaN nanowire heterostructures (1.58-1.95 µm) [28], by using smaller nanodisks, we were able to tune the absorption of our heterostructures to the telecommunication wavelength and improve the FWHM from ~400 meV [22] to 200 meV. This improvement is due to the fact that large quantum wells (>3 nm) require high doping levels (>$1\times10^{20}$cm$^{-3}$) to absorb in the telecommunication wavelength, which leads to the broadening of transitions due to many-body effects.

In NS2, NG2 samples, we observe a blueshift with respect to NS1, NG1, which can be attributed to many-body effects. ISB transitions from the largest doped nanowires (NG3 and NS3) of $3\times10^{20}$ cm$^{-3}$ could not be clearly discerned, probably due to the large broadening of the transition. The large, inhomogeneous broadening of the absorption peaks is due to thickness/diameter variations in the nanowire ensemble. This prevents resolving monolayer thickness fluctuations in the axial direction, similar to observations in GaN/AlN quantum dots grown by the Stranski-Krastanov method [41].

3D calculations of the nanowire quantum confined ground and excited states were performed using nextnano3 using k.p model. For a doping concentration $N_D = 3\times10^{19}$cm$^{-3}$, the s-p$_z$ transition is calculated at 0.729 eV, which is indicated by arrows in figures 5 (b) and (d). Experimentally, even though the observed blueshift with increasing doping can be attributed to many-body effects, when considering 3D calculations, many-body effects due to doping could not be taken into account to calculate the ISB transitions for samples with higher doping levels. An approach similar to the planar samples taking into account the 3D calculations is beyond the scope of this work, but the results from the 1D calculations can be used as a guideline to the expected blueshift.

## 4. Conclusion

In conclusion, we synthesized GaN/AlN nanowire heterostructures and planar quantum wells with similar periodicity using PAMBE, varying dopant type (Si and Ge) and dopant concentration in the GaN sections. The choice of quantum well/nanodisk size was made to fit for ISB absorption in the SWIR region. With increasing dopant concentration, planar heterostructures exhibited a blueshift of the PL peak emission due to the screening of the internal electric fields by free carriers. For both nanowire and planar geometries, the PL peak energy, linewidth and decay time constant were not influenced by the choice of dopant (Si or Ge). However, we have verified that nanowire heterostructures consistently present longer PL decay times than their planar counterparts with identical nanodisk/well size and doping level. This experimental evidence confirms the existence of an in-plane piezoelectric field in nanowires associated to the sheer component of the strain tensor, leading to lateral electron-hole separation.

Regarding ISB transitions, we present planar Ge-doped heterostructures exhibiting absorption characteristics associated to the transition from first to second electronic level in the quantum well covering 1.45 µm to 1.75 µm, which was identical to the Si-doped heterostructures used as a reference. To the best of our knowledge, these results constitute the first systematic study of ISB transitions in Ge-doped planar GaN-based structures. Blueshift and broadening of the ISB transitions associated to many-body effects, including the exchange interaction and depolarization shift, have been calculated and found to fit the experimental data.

In nanowire heterostructures, we reported the first observation of ISB absorption in GaN:Si/AlN heterostructures with varying doping levels, which also lead to an improved FWHM of 200 meV for the ISB absorption at 1.55 µm. The inhomogeneous broadening was associated to geometry fluctuations, regardless of doping. Si-doped and Ge-doped nanowires behaved identical, indicating that the choice of dopant is not a hindrance for observing ISB transitions in nanowires. Based on this study, we conclude that both Si- and Ge-doped nanowires are potentially suitable for the fabrication of GaN/AlN heterostructures for the study of ISB optoelectronic phenomena.

**Acknowledgements.** The authors acknowledge technical support by Y. Genuist and Y. Curé, This work is supported by the EU ERC-StG "TeraGaN" (#278428) project, and by the French National Research Agency via the GaNEX program (ANR-11-LABX-0014) and ANR-COSMOS (ANR-12-JS10-0002). We benefited from access to the technological platform NanoCarac of CEA-Minatech. The LANEF framework (ANR-10-LABX-51-01) is acknowledged for its support with mutualized infrastructure

**FIGURE CAPTIONS**

**Figure 1.** (a) Tilted view (45º) of sample NG1 using SEM. (b) Schematic representation of the nanowire heterostructures under study, (c, d) HAADF-STEM images of a nanowire from NG1 at two different magnifications. Dark/bright contrast corresponds to AlN/GaN. The direction of observation was [1-100].

**Figure 2.** Low-temperature PL spectra of (a) Si-doped planar heterostructures (b) Si-doped nanowire heterostructures (c) Ge-doped planar heterostructures (d) Ge-doped nanowire heterostructures. Spectra are normalized to their maximum and vertically shifted for clarity. The GaN bandgap is indicated by a dotted vertical line.

**Figure 3.** Band-to-band $e_1$-$h_1$ energy as a function of the quantum well/nanodisk width for (a) planar heterostructures, and (b) nanowire heterostructures. The solid line indicates theoretical calculations. The data points with error bars are experimental PL peak position values. The horizontal error bar indicates the error in the determination of the well width, and the vertical error bar marks the PL peak wavelengths obtained for the samples doped with the same Si and Ge concentration (precise values compiled in Table I). Squares represent data from ref. [28], with the doping levels indicated in the legend.

**Figure 4.** Decay time constant extracted from time-resolved PL as a function of doping concentration. The vertical error bar indicates the values obtained for Si- and Ge-doped samples of the same dopant concentration. Inset: Normalized decay of the PL intensity as a function of time for NG1 and PG1. The laser pulse arrives at time t = 0.

**Figure 5.** Room-temperature absorption spectra for TM-polarized light measured in (a) Si-doped planar heterostructures, (b) Si-doped nanowire heterostructures, (c) Ge-doped planar heterostructures, and (d) Ge-doped nanowire heterostructures. Spectra are normalized to their maximum and vertically shifted for clarity. Theoretical values

combining 1D nextnano$^3$ and many-body calculations are indicated as arrows in (a), (c). Theoretical values from 3D nextnano3 calculations are indicated as arrows in (b), (d).

**Table I.** Structural and optical characteristics of planar and nanowire GaN/AlN heterostructures: Heterostructure geometry (planar/nanowire); Doping element (Si/Ge); Dopant concentration ($N_D$); Period of the heterostructures measured using HR-XRD (GaN+AlN width); Low temperature PL peak position; PL decay time constant extracted from time-resolved PL measurements as described in the text.

| Sample | Geometry | Dopant | $N_D$ (cm$^{-3}$) | Period (nm) | PL peak position (nm) | PL decay constant (ns) |
|---|---|---|---|---|---|---|
| PS0 | Planar | Si | $1\times10^{18}$ | 4.8±0.1 | 366 | 11.6 |
| PG0 | Planar | Ge | $1\times10^{18}$ | 4.8±0.1 | 369 | 7.9 |
| PS1 | Planar | Si | $3\times10^{19}$ | 4.8±0.1 | 368 | 7.3 |
| PG1 | Planar | Ge | $3\times10^{19}$ | 4.8±0.1 | 370 | 11.6 |
| PS2 | Planar | Si | $1\times10^{20}$ | 4.8±0.1 | 359 | 3.2 |
| PG2 | Planar | Ge | $1\times10^{20}$ | 4.7±0.1 | 358 | 5.6 |
| PS3 | Planar | Si | $3\times10^{20}$ | 4.3±0.1 | 353 | 3.5 |
| PG3 | Planar | Ge | $3\times10^{20}$ | 4.2±0.1 | 341 | 2.5 |
| N0 | NW | - | - | 4.4±0.3 | 345 | 14.0 |
| NG1 | NW | Ge | $3\times10^{19}$ | 4.6±0.2 | 336 | 16.0 |
| NS1 | NW | Si | $3\times10^{19}$ | 4.6±0.2 | 339 | 15.3 |
| NG2 | NW | Ge | $1\times10^{20}$ | 4.6±0.2 | 336 | 10.2 |
| NS2 | NW | S | $1\times10^{20}$ | 4.3±0.2 | 338 | 8.9 |
| NG3 | NW | Ge | $3\times10^{20}$ | 4.1±0.2 | 336 | 3.3 |
| NS3 | NW | Si | $3\times10^{20}$ | 4.6±0.2 | 337 | 5.4 |

**Table II.** Numerical calculation of the many-body effects involved in the blueshift of the ISB absorption: Dopant concentration ($N_D$); Theoretical e$_2$-e$_1$ transition energy for low doping levels ($E_{21}$); Shift due to exchange interaction ($E_{ex}$); Coefficient of depolarization shift ($\alpha$); Corrected e$_2$-e$_1$ transition energy using equation (1) ($E'_{21}$); Experimentally observed peak energy of the ISB absorption.

| Sample | $N_D$ (cm$^{-3}$) | $E_{21}$ (meV) | $E_{ex}$ (meV) | $\alpha$ | $E'_{21}$ (meV) | ISB transition (meV) |
|---|---|---|---|---|---|---|
| PS1, PG1 | 3×10$^{19}$ | 729 | 0.09 | 3×10$^{-5}$ | 729 | 716 |
| PS2, PG2 | 1×10$^{20}$ | 729 | 3 | 8×10$^{-4}$ | 732 | 811, 778 |
| PS3, PG3 | 3×10$^{20}$ | 787 | 30 | 4×10$^{-4}$ | 817 | 873, 858 |

**Figure 1**

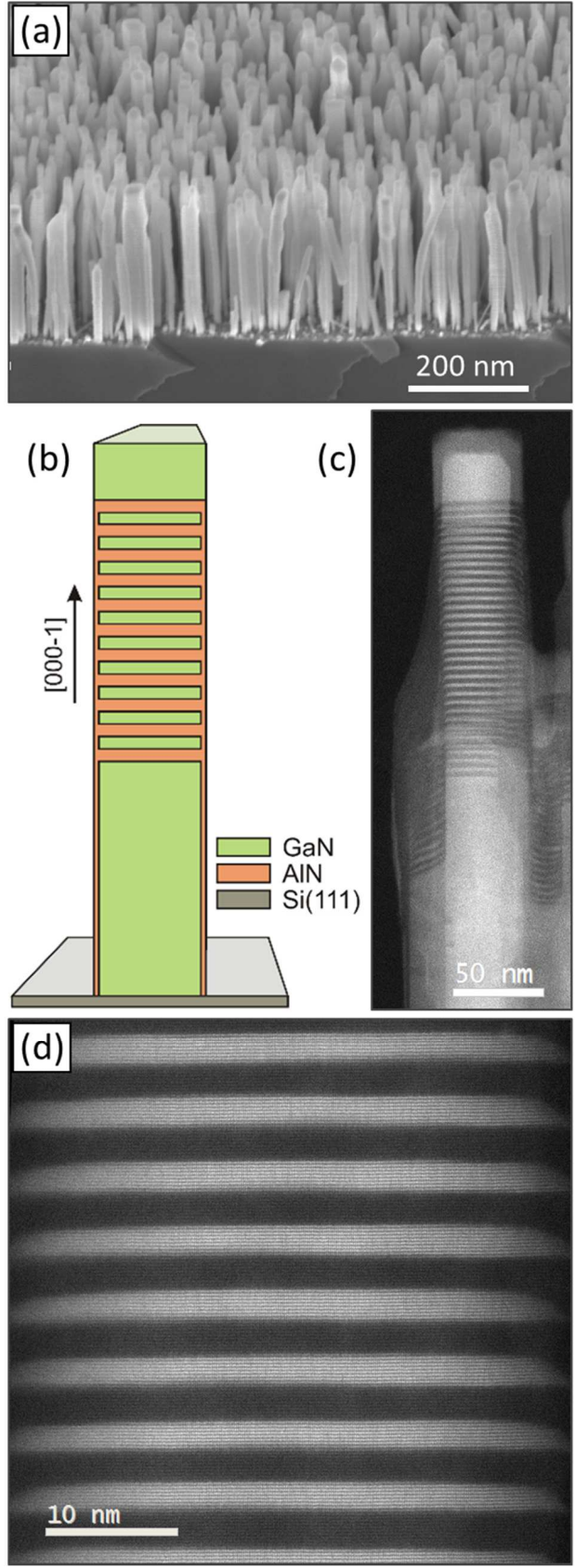

**Figure 2**

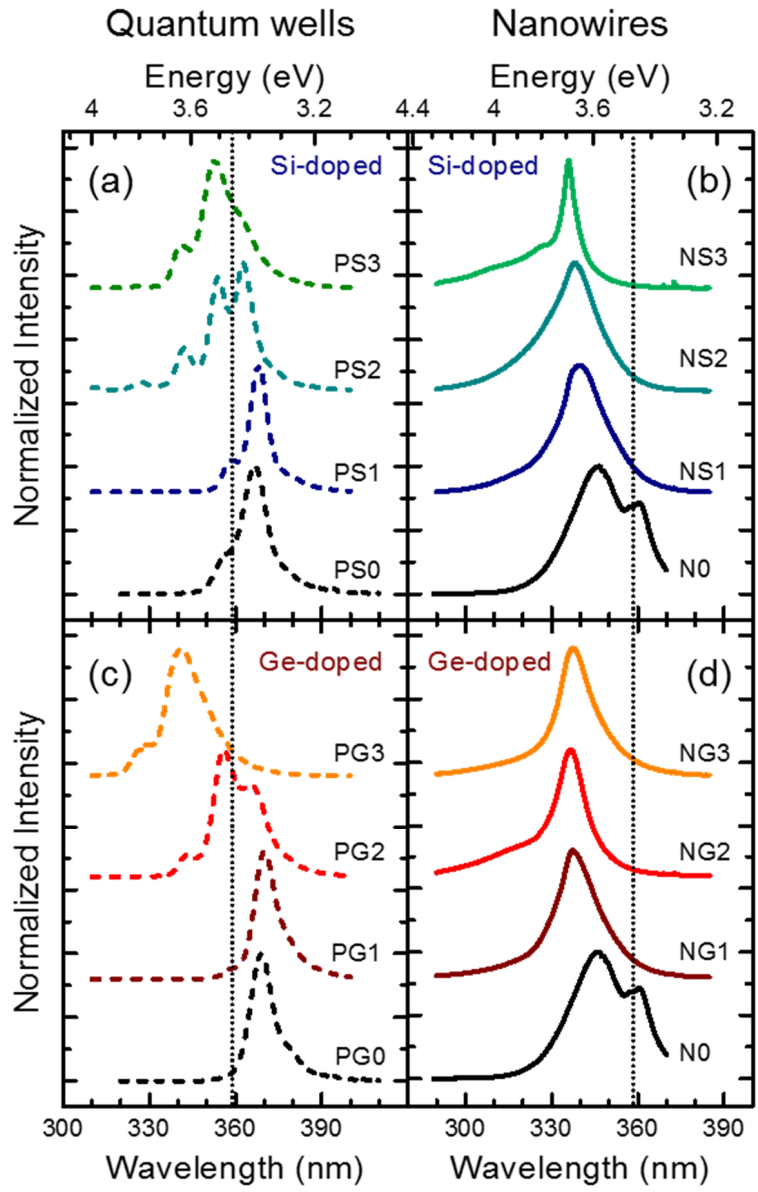

**Figure 3**

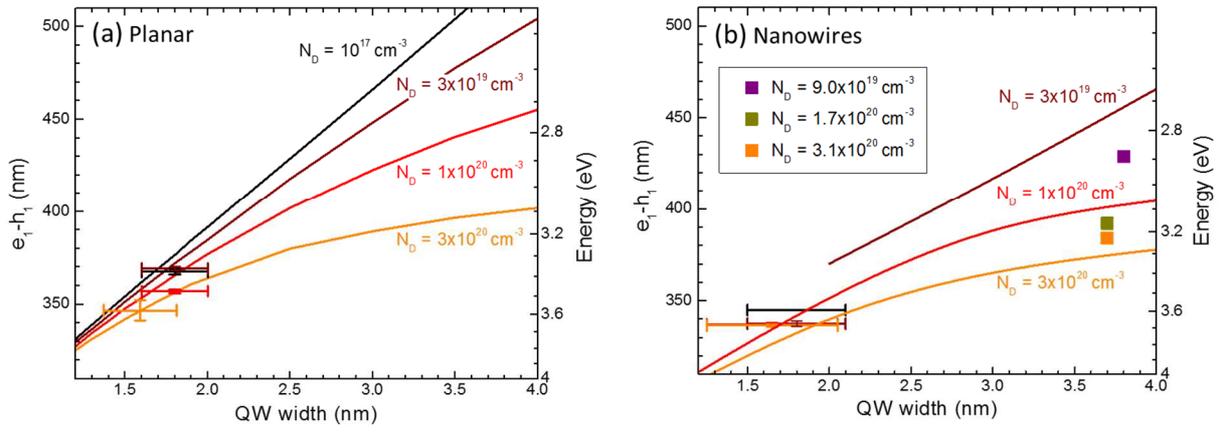

**Figure 4**

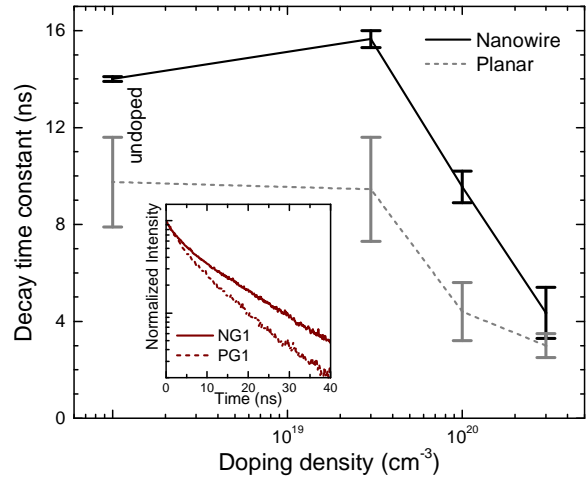

**Figure 5**

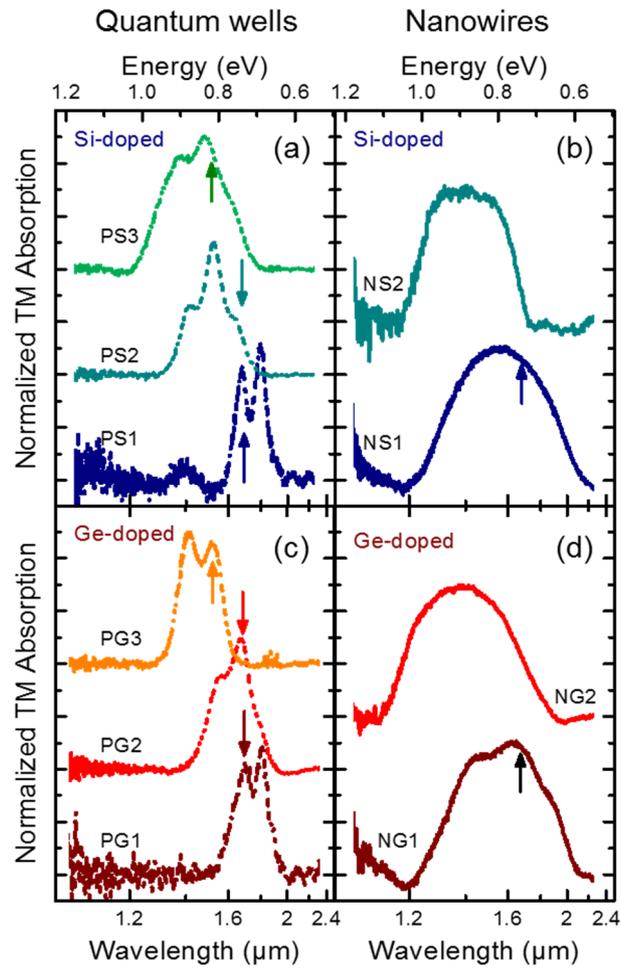